\documentclass[prd,aps,amssymb,nofootinbib,showpacs,preprint]{revtex4}
\usepackage{amsmath,amsfonts}
\usepackage{hyperref}

\newcommand{\mds}{m_{ \mbox{\tiny \sl DS}}}
\newcommand{\beq}{\begin{equation}}
\newcommand{\eeq}{\end{equation}}
\newcommand{\bear}{\begin{eqnarray}}
\newcommand{\ear}{\end{eqnarray}}
\newcommand{\earn}{\nonumber \end{eqnarray}}

\newcommand{\nn}{\nonumber \\}

\begin{document}

\title{Semiclassical long throats of the wormholes}

\author{A. Popov}
\address{Kazan Federal University, 18 Kremlyovskaya St., Kazan 420008, Russia\\
apopov@kpfu.ru}

\begin{abstract}
The vacuum polarization of the quantized scalar field with non-conformal
coupling $\xi$ on the long throat of a wormhole background is calculated.
It is shown that the stress-energy tensor of vacuum fluctuations in considered
spacetime is determined by the local geometry of spacetime only.
The self-consistent solutions of the semiclassical Einstein field
equations describing a long throat of a traversable wormhole are obtained.
\end{abstract}

\maketitle

\section{Introduction}

By definition a wormhole is a bridge connecting two
asymptotically flat regions. Usually such construction is a
classical object and should satisfy to Einstein equations. The
topology of the 4D wormholes is the topology of direct product
of the Minkowski plane and a unit sphere. Static traversable
wormholes could be threaded by ''exotic matter'' that violates
certain energy conditions at least at the throat \cite{MTY}. As an
example of such a matter one can consider the vacuum of quantized
fields. This approach gives the possibility to consider the
wormhole metric as the self-consistent solution of the
semiclassical theory of gravity. In the realm of this theory the
vacuum fluctuations of the quantized fields are the source of
spacetime curvature\footnote{throughout we use units such that
$c= \hbar =G=1$.}
     \beq \label{1}
     G^{\mu}_{\nu} = 8\pi \langle T^{\mu}_{\nu}
     \rangle_{ren}.
     \eeq
The main difficulty in the theory of semiclassical gravity is that
the effects of the quantized gravitational field are ignored. The
popular solution of this problem is to justify ignoring the
gravitational contribution by working in the limit of a large
number of fields, in which the gravitational contribution is
negligible. Another problem is that the vacuum polarization
effects are determined by the topological and geometrical
properties of spacetime as a whole or by the choice of quantum
state in which the expectation values are taken. It means that
calculating of the functional dependence of $\langle T^{\mu}_{\nu}
\rangle_{ren}$ on the metric tensor in an arbitrary spacetime
presents formidable difficulty. Only in some spacetimes with high
degrees of symmetry for the conformally invariant fields $\langle
T_{\mu \nu} \rangle_{ren}$ can be computed and equations (\ref{1})
can be solved exactly \cite{KSS}. Let us stress that the single
parameter of length dimensionality in such a problem is
the Planck length ${l_{ \mbox{\tiny \sl Pl}}}$. This implies that
the characteristic scale $l$ of the spacetime curvature (which
correspond to the solution of equations (\ref{1})) can differ from
${l_{ \mbox{\tiny \sl Pl}}}$ only if there is a large
dimensionless parameter. As an example of such a parameter one can
consider a number of fields the polarization of which is a source
of spacetime curvature\footnote{Here and below it is
assumed, of course, that the characteristic scale of change of the
background gravitational field is sufficiently greater than ${l_{
\mbox{\tiny \sl Pl}}}$ so that the very notion of a classical
spacetime still has some meaning.}. In some cases $\langle T_{\mu
\nu} \rangle_{ren}$ is determined by the local properties of a
spacetime and it is possible to calculate the functional
dependence of the renormalized expression for the vacuum
expectation value of the stress-energy tensor operator of the
quantized fields on the metric tensor approximately. One of the
most widely known examples of such a situation is the case of a
very massive field. In this case $\langle T^{\mu}_{\nu}
\rangle_{ren}$ can be expanded in terms of powers of the small
parameter
      \beq \label{ml}
      \frac1{m l} \ll 1,
      \eeq
where $m$ is the mass of the quantized field and $l$ is the
characteristic scale of the spacetime curvature \cite{MatK,AHS}.
Using the first nonvanishing term of this expansion for minimally
or conformally coupled scalar field Taylor, Hiscock and Anderson
\cite{THA} have showed that the equations (\ref{1}) have no wormhole
solution for some class of static spherically symmetric
spacetimes. Let us stress that in this case the existence of an
additional parameter of the length dimensionality $1/m$ does not
increase the characteristic scale of the spacetime curvature which
is described by the solution of equations (\ref{1})\footnote{The
characteristic scale of the components $G^{\mu}_{\nu}$ on the
left-hand side of equations (\ref{1}) is $1/l^2$, on the
right-hand side - ${l_{ \mbox{\tiny \sl Pl}}}^2/(m^2 l^6)$.}.

It is necessary to note that different quantum field theory
constraints on the parameters of traversable wormholes are known
\cite{FRNZK}. As a rule these constraints were derived for a
massless scalar field.

The purpose of this paper is to examine whether the vacuum
fluctuations of quantized fields in the one-loop approximation in
Einstein's theory can create traversable Lorentzian wormholes. One
of the possible ways to solve this problem is to use the
analytical approximation for the expectation value of the
stress-energy tensor operator of the quantized matter fields in
curved spacetimes \cite{AHS,Num}. This approach was realized in
the works \cite{HPS}. The problem of such an approach is the
uncertainty of the applicability limits of such approximations. As
it was noted by Khatsymovsky \cite{V} in the spacetime that is a
direct product of the Minkowski plane and a two-dimensional sphere
of a fixed radius (the topologies of this spacetime and wormhole
spacetime coincide) these approximations are not applicable.
Another way to obtain the wormhole solution of equations (\ref{1})
is to use the model of short-throat flat-space wormhole \cite{KS}
(see also \cite{K}). This model represents two identical copies of
Minkowski spacetime with spherical regions excised from each copy
and so that points of these regions are identical. One can
consider this model as the first approximation of real situation
if there is a small parameter $L/r$, where $L$ is the length of
the throat and $r$ is the radius of the throat. At the present
time only the full vacuum energy of quantized scalar field have
been calculated in this spacetime. Nevertheless it gives a
possibility to make some evaluations of the radius of a wormhole
throat. A completely opposite model is the model of a long-throat
wormhole. Local approximations for $\langle T^{\mu}_{\nu} \rangle$
in the throat of static spherically symmetric long-throat wormhole
were obtained in \cite{Khat} for massless fields of spin 1 and
1/2. The results of these works were obtained by the WKB method
and the small parameter of these approximations is the ratio of
the throat radius to the length of the throat. This result gives
the possibility to reply to the question: can the throat of a
long-throat wormhole be created by the vacuum fluctuations of
quantum fields?

In this paper, the results of the work \cite{Popov} are used to
evaluate the local approximation of the renormalized expression
for the vacuum expectation value of the stress-energy tensor
operator of the quantized scalar field in the throat of static
spherically symmetric long-throat wormhole spacetime (Sec.2). In
Sec.3 the solution of semiclassical Einstein field
equations describing the long throat of a traversable Lorentzian
wormhole are given. Sec.4 summarizes the contents.

\section{Stress-energy of a quantized scalar field in spacetime of
static spherically symmetric long  throat wormhole}

The line element of a static spherically symmetric wormhole
spacetime can be written as
     \beq \label{metric}
     ds^2 = -f dt^2 + d \rho^2 + r^2 (d \theta^2 +\sin^2 \theta d
     \phi^2).
     \eeq
where $ - \infty < \rho < \infty$ and $f=f(\rho), \ r=r(\rho)$.

The vacuum polarization effects of a quantized field can be
described approximately \cite{Khat, Popov} in the region of
variation of $\rho$ where the functions $f(\rho), r(\rho)$ change
slowly. One can consider such a region as a long throat of a
wormhole. It is necessary to note that the notion of throat's
length is not well-defined. To discuss this notion let us consider
the model of wormhole
     \bear
     ds^2 = - dt^2 + d \rho^2 &+& \left[\rho \tanh
     \left({\rho}/{\lambda} \right)
     \right. \nn && \left.
     +r_0 \right]^2 (d \theta^2 +\sin^2 \theta d \phi^2).
     \ear
In this model the parameter $r_0$ describes the radius of the
throat. In the region $|\rho| \gg \lambda$ the spacetime is
asymptotically flat. A curved region $|\rho| \lesssim \lambda$ can be
called the throat of the wormhole and $\lambda$ characterizes the
length of this throat (see, also, \cite{Sush}).
In a more general case the length of
the throat can be defined as a scale characterizing the variation
of the metric function $r^2(\rho)$ at the throat provided that
this scale is much more than the radius of the throat.
Additionally one should assume that $f(\rho)$ is changed slowly.

The renormalized expressions for the vacuum expectation value of
the stress-energy tensor operator of the quantized scalar field in
the region of variation of $\rho$ where the functions $f(\rho)$
and $r(\rho)$ change slowly can be expanded (see \cite{Popov}) in
terms of powers of the small parameter
     \beq \label{lwkb}
     L_{\star}(\rho) /L(\rho) \ll 1,
     \eeq
where
      \beq
      L_{\star}(\rho) =\left[ m^2+ \frac{2\xi}{r^2} \right]^{-1/2},
      \eeq
$m$ is the mass of the scalar field, $\xi$ is its coupling to the
scalar curvature and $L(\rho)$ is a scale of variation of the
metric functions:
       \bear \label{Lm}
       \frac{1}{L(\rho)}&& = \max \left \{ \left| \frac{r'}{r}  \right|, \
       \left| \frac{f'}{f}  \right|, \
       \sqrt{\left|\xi \right|} \left| \frac{r'}{r} \right|, \
       \sqrt{\left|\xi \right|} \left| \frac{f'}{f} \right|, \
       \right. \nn && \left.
       \left| \frac{r''}{r}  \right|^{1/2}, \
       \left| \frac{f''}{f}  \right|^{1/2}, \
       \sqrt{\left|\xi \right|} \left| \frac{r''}{r}  \right|^{1/2}, \
       \dots  \right \} .
       \ear
The zeroth-order terms (with respect to a small parameter
$L_{\star} /L$) of the renormalized expression for the vacuum
expectation value of the stress-energy tensor operator of the
quantized scalar field in such a spacetime is given by (see
\cite{Popov})
     \bear \label{T_0_tt}
     &&\left< T^{t}_{t} \right>_{ren}=\left< T^{\rho}_{\rho} \right>_{ren}=
     \frac{1}{4 \pi^2 r^4}\left\{\frac{m^2 r^2}{8 }\left( \xi-\frac18 \right)
     \right.\nn &&
     + \frac{79}{7680}-\frac{11}{96}\xi+\frac38 \xi^2
     + \left[-\frac{m^4 r^4}8+\frac{m^2 r^2}{2}\left(\frac16-\xi \right)
     \right.\nn && \left.
     -\frac{1}{60}+\frac{1}{6}\xi-\frac12 \xi^2
     \right]\ln\sqrt{\frac{\mu^2}{\mds^2r^2}}
     + \left[{2 m^2 r^2}\left(\xi-\frac18 \right)
     \right. \nn && \left. \left.
     + \frac{m^4 r^4}2
     +2\left( \xi-\frac18 \right)^2 \right]
     \left[I_1(\mu)-I_2(\mu)\right]  \right\},
     \ear
     \bear \label{T_0_thth}
     &&\left< T^{\theta}_{\theta} \right>_{ren}=
     \left< T^{\varphi }_{\varphi} \right>_{ren}=
     \frac{1}{4 \pi^2 r^4}\left\{\frac{m^2 r^2}{8}\left( \xi-\frac18 \right)
     \right.\nn &&
     + \left[-\frac{m^4 r^4}8
     -\frac1{8 } \left(\xi-\frac18 \right)^2
     +\frac{1}{60}-\frac{1}{6}\xi
     \right. \nn && \left.
     +\frac12 \xi^2 \right]\ln\sqrt{\frac{\mu^2}{\mds^2r^2}}
     + \left[{m^2 r^2}\left(\frac18-\xi \right)
     \right. \nn && \left.
     -2\left( \xi-\frac18 \right)^2 \right]I_1(\mu)
     + \left[\frac{m^4 r^4}{2}
     \right. \nn && \left. \left.
     +{2m^2 r^2}\left(\xi-\frac18 \right)
     +2\left( \xi-\frac18 \right)^2
      \right] I_2(\mu)  \right\},
     \ear
     \beq \label{T_0_mn}
     \left< T^{\mu}_{\nu} \right>_{ren}=0,
     \ \mu\neq \nu ,
     \eeq
where
     \beq
     \mu^2=m^2r^2+2\xi-1/4 > 0,
     \eeq
     \bear
     &&I_1(\mu)=\int \nolimits_{0}^{\infty}\frac{x\ln|1-x^2|}
     {1+e^{2\pi |\mu| x}}dx,
     \nn &&
     I_2(\mu)=\int \nolimits_{0}^{\infty}\frac{x^{3}\ln|1-x^2|}
     {1+e^{2\pi |\mu| x}}dx,
     \ear
$\mds$ is equal to the mass $m$ of the field for a massive scalar
field. For a massless scalar field it is an arbitrary parameter
due to the infrared cutoff in renormalization counterterms for
$\left< T^{\mu}_{\nu} \right>$. A particular choice of the value
of $\mds$ corresponds to a finite renormalization of the
coefficients of terms in the gravitational Lagrangian and must be
fixed by experiment or observation. Note that the renormalized
expectation values of the stress-energy tensor components
(\ref{T_0_tt}-\ref{T_0_mn}) are exact if $f(\rho)=const$ and
$r(\rho)=const$. Note also that $\left< T^{\mu}_{\nu}
\right>_{ren}$ is conserved
     \beq
     \left< T^{\mu}_{\nu} \right>_{ren; \mu}=0
     \eeq
and, for the conformally invariant field, has a trace equal to the
known trace anomaly \cite{BD}
     \bear
     &&\left< T^{\mu}_{\mu} \right>_{ren}= \frac{-1}{2880 \pi^2}
     \left[C_{\alpha\beta\gamma\delta}C^{\alpha\beta\gamma\delta}
     +R_{\alpha\beta} R^{\alpha\beta}
     \right. \nn && \left.
     -\frac13 R^2 -  \Box R
     \right] = -\frac{1}{4 \pi^2 r^4} \frac{1}{360}.
     \ear

The expressions (\ref{T_0_tt}) and (\ref{T_0_thth}) may be
simplified in the cases of massless field.
In particular, if
     \beq
     m = 0, \xi=1/6, L^2 \gg \frac{r^2}{2 \xi},
     \eeq
then
     \bear \label{tt}
     \left< T^{t}_{t} \right>_{ren}=\left< T^{\rho}_{\rho} \right>_{ren} \simeq &&
     \frac{1}{4 \pi^2 r^4} \left[ \frac{}{}0.00310
     \right. \nn && \left.
     +\frac1{720} \ln{(\mds^2 r^2)}
      \right],
      \ear
      \bear \label{thth}
      \left< T^{\theta}_{\theta} \right>_{ren}=
      \left< T^{\varphi }_{\varphi} \right>_{ren}
      \simeq &&
     \frac{1}{4 \pi^2 r^4}  \left[ \frac{}{} -0.00171
     \right. \nn && \left.
     -\frac1{720} \ln{(\mds^2 r^2)}
      \right].
      \ear
In the case of a very massive field
     \beq
     m^2 r^2 \gg \left|\, 2 \xi\right|, \ \mu^2 \gg 1,
      \  L^2 \gg \frac{r^2}{2 \xi}
     \eeq
the expressions for $I_n(\mu)$ and
$\ln(\sqrt{{\mu^2}/{\mds^2r^2}})$ can be expanded in terms of
powers of $(2 \xi-1/4)/(m^2 r^2)$. As a result the expressions
(\ref{T_0_tt}-\ref{T_0_mn}) can be written in the form
       \bear \label{gg}
       &&\left< T^{\mu}_{\nu} \right>_{ren}= \frac1{4\pi^2r^4}
       \left[\frac1{m^2r^2}\left(
       \frac{\xi^3}{6}-\frac{\xi^2}{12} +\frac{\xi}{60}-\frac1{630} \right)
       \right. \nn && \left.
       +O\left(\frac{\left(2 \xi - 1/4\right)^2}{m^4 r^8} \right)
       \right] \left( \begin{array}{rrrr}\!-1&0&0&0\\0&\ \!-1&0&0\\
       0&0&\ \ 2&0\\0&0&0&\ \ 2 \end{array} \right).
        \ear
These expressions can be obtained directly from the correspondent
terms of the DeWitt-Schwinger approximation if we assume that
$f=const$ and $r^2=const$.

\section{Solutions}

As it was mentioned above the right hand side of equations (\ref{1})
may be expanded in terms of powers of the small parameter
$L_{\star}/L$ in the region of variation of $\rho$ where the
metric functions $f(\rho)$ and $r(\rho)$ change slowly. This
implies that these equations may be solved iteratively. And to
obtain the zeroth-order solution of equations (\ref{1}) one must
omit the terms of order $1/L^2$ in left hand side of these
equations the same way as the corresponded terms in right hand
side was omitted (in expressions for $\left< T^{\mu}_{\nu}
\right>_{ren}$ (\ref{T_0_tt},\ref{T_0_thth}))
    \bear \label{Gmn}
    G^t_t=-\frac1{r^2}&-&\frac{(r^2)'^{2}}{4 r^4}
    +\frac{(r^2)''}{r^2} =-\frac1{r^2} +O\left(1/{L^2} \right),
    \nn
    G^{\rho}_{\rho}=-\frac1{r^2}&+&\frac{(r^2)'^{2}}{4 r^4}
    +\frac{f'(r^2)'}{2 f r^2} =-\frac1{r^2} +O\left(1/{L^2}
    \right), \nn
    \ G^{\theta}_{\theta} = G^{\varphi}_{\varphi} &=&  \frac{(r^2)''}{2
    r^2} +\frac{f''}{2 f} +\frac{f' (r^2)'}{4 f r^2} -\frac{(r^2)'^{2}}{4
    r^4}
    \nn
    && - \frac{f''}{4 f^2} =  O\left(1/{L^2} \right).
    \ear
The zeroth-order solution is valid in the region of variation of
$\rho$ where $L_{\star}(\rho)/L(\rho) \ll 1$. As it was mentioned
above one can consider such a region as a long throat of a
wormhole.

The principal problem of the solution of equations (\ref{1}) in
considered case is that we have two equations on only one unknown
variable $r^2$. This problem can be overcome by the insertion into
consideration the classical electrostatic field. The stress-energy
tensor of such a field created by the charge $Q$ in coordinates
(\ref{metric}) is
    \beq
 T^{\mu}_{\nu}=\frac{Q^2}{8 \pi r^4} \ diag\left( -1,-1,1,1 \right).
    \eeq
Thus the parameter $Q$ plays a role of an additional variable.

Now let us consider the case in which the right hand side of
equations (\ref{1}) is determined by this
electrostatic field and two quantized scalar fields: conformally
invariant field ($\xi_1=1/6, m_1=0$) and very massive field
($\xi_2 \equiv \xi$ - arbitrary constant, $m_2 \equiv m $,
$\mu_2^2 = m^2 r^2 + 2 \xi - 1/4 \gg 1$,  $m^2 r^2 \gg |2 \xi
-1/4|$). For this case the Einstein's equations have the form
     \bear \label{15}
      -\frac1{8 \pi r^2} \simeq \frac1{4 \pi^2 r^4} \left[ 0.00310 +
     \frac1{720} \ln{(\mds^2 r^2)} \right. \nn  \left.
     +\frac1{m^2 r^2} \left(
     -\frac{\xi^3}{6} +\frac{\xi^2}{12} -\frac{\xi}{60} +\frac1{630}
     \right) \right] -\frac{Q^2}{8 \pi r^4},
     \ear
     \bear \label{16}
     0  \simeq  \frac1{4 \pi^2 r^4} \left[ -0.00171 -
     \frac1{720} \ln{(\mds^2 r^2)} \right. \nn \left.
     +\frac1{m^2 r^2} \left(
     \frac{\xi^3}{3} -\frac{\xi^2}{6} +\frac{\xi}{30} -\frac1{315}
     \right) \right] +\frac{Q^2}{8 \pi r^4}.
     \ear
The consequence of these equations is
      \beq
      r^4 +\frac{r^2}{360\pi} +\frac1{\pi m^2} \left(
      \frac{\xi^3}{3} -\frac{\xi^2}{6} +\frac{\xi}{30} -\frac1{315}
      \right) \simeq 0.
      \eeq
The solution of this equation which satisfies the condition
      \beq \label{con}
      r^2 \gg 1/ (720 \pi)
      \eeq
is
      \beq \label{r}
      r^2 \simeq \sqrt{-\frac1{\pi m^2 } \left( \frac{\xi^3}{3}
      -\frac{\xi^2}{6} +\frac{\xi}{30} -\frac1{315} \right) }.
      \eeq
The second independent equation of system (\ref{15},\ref{16})
imposes a constraint on $Q^2$
      \bear \label{Q}
      &&Q^2 \simeq  \frac{2}{\pi}  \left\{
      0.00171+\frac1{720} \ln \left[\frac{\mds^2}{m \sqrt{\pi}}
      \left( -\frac{\xi^3}{3} +\frac{\xi^2}{6}
      \right. \right. \right.  \nn && \left. \left. \left.
      -\frac{\xi}{30} +\frac{1}{315} \right)^{1/2}\right]
      +\frac{\sqrt{\pi}}{m} \left( -\frac{\xi^3}{3}
      +\frac{\xi^2}{6} -\frac{\xi}{30}
      \right. \right.  \nn && \left. \left.
      +\frac{1}{315} \right)^{1/2}
      \right\}.
      \ear
If we take into account (\ref{con}) and $m^2 r^2 \gg |2 \xi -1/4|$,
the conditions of validity of this solution can be written as follows
      \bear
      &&- \frac{\displaystyle \pi \left({2\xi - 1/4}\right)^2}
      {\displaystyle \left(\frac{\xi^3}{3} -\frac{\xi^2}{6}
      +\frac{\xi}{30} -\frac1{315} \right) } \ll m^2 \ll
      \nn &&
       -518400 \, \pi \left( \frac{\xi^3}{3}
      -\frac{\xi^2}{6} +\frac{\xi}{30} -\frac1{315}
      \right).
      \ear
It is necessary to remember that the region of
validity of the semiclassical theory of gravity is determined by
the condition $r \gg 1$. All these conditions are valid for $\xi<0, \
|\xi| \gg 1$.
A particular solution of system (\ref{15},\ref{16}) is
      \beq
      \xi=-10^{4}, \ m^2 = 10^{3}, \
      \ r \simeq 101.49 .
      \eeq
Let us note that the stress-energy of the fields considered here
have the needed "exotic" properties (in the sense of Morris and
Thorne \cite{MTY}) to support the long throat of a wormhole:
\bear
&&p_{r}=-\epsilon=\frac1{4 \pi^2 r^4} \left[ 0.00310 + \frac1{720} \ln{(\mds^2 r^2)} \right. \nn && \left.
     +\frac1{m^2 r^2} \left( -\frac{\xi^3}{6} +\frac{\xi^2}{12} -\frac{\xi}{60} +\frac1{630}
     \right) \right]
     \nn && -\frac{Q^2}{8 \pi r^4} < 0,
\ear
where $p_{r}$ is the radial pressure, $\epsilon$ is the energy
density and $r, Q$ are determined by expressions (\ref{r},\ref{Q}).

\section{Conclusions}

In this paper we succeed in finding the solution in semiclassical theory of gravity which describes the long throat of the wormhole. Such objects are created by the electrostatic field and the vacuum fluctuations of quantized scalar fields. The stress-energy tensor of these fluctuations in considered spacetime is determined by the local geometry of
spacetime only. The geometry of spacetime far from the throat is not described by the obtained  zeroth-order (with respect to the small parameter
$L_{\star}/L$) solution. The obvious defect of such solutions is as follows: the value of $|\xi|$ which corresponds to the large (with
respect to the Planck length) value of throat radius is also large (with respect to 1). The latter seems unlikely. Nevertheless in
the "large $N$" case, in which the number of matter fields is large, the radius of throat $r$ is proportional to $\sqrt{N}$
($N^{1/4}$ for the very massive fields) and the value of $r$ which satisfy to the condition $r \gg 1$ (i.e. much more than the Planck
length) can be obtained for the value of $|\xi|$ lesser than one considered above.

{\bf Acknowledgments}\\
\noindent
 This work was supported in part by grant 13-02-00757
from the Russian Foundation for Basic Research.

\end{document}